\documentclass[a4paper,11pt]{article}
\usepackage{jheppub} % for details on the use of the package, please see the JINST-author-manual
\usepackage{lineno}
\usepackage{braket}
\usepackage{amsmath}
\usepackage{accents}
\newcommand{\B}{\boldsymbol}
\title{\boldmath Probing the 3+1 neutrino model in the SHiP experiment}

\author[a]{Ki-Young Choi,}
\author[a]{Yu Seon Jeong,}
\author[c]{Sung Hyun Kim,}
\author[d]{Yeong Gyun Kim,}
\author[c]{Kang Young Lee,}
\author[e]{Kyong Sei Lee,}
\author[c]{Byung Do Park,}
\author[c]{Jong Yoon Sohn,}
\author[a]{Seong Moon Yoo}
\author[c]{and Chun Sil Yoon}

\affiliation[a]{Department of Physics and Institute of Basic Science, Sungkyunkwan University, Suwon 16419, Korea}
\affiliation[c]{Department of Physics Education \& Research Institute of Natural Science,
Gyeongsang National University, Jinju 52828, Korea}
\affiliation[d]{Department of Science Education, Gwangju National University of Education, Gwangju 61204, Korea}
\affiliation[e]{Center for Extreme Nuclear Matters, Korea University, Seoul 02841, Korea}

\emailAdd{kiyoungchoi@skku.edu}
\emailAdd{castledoor@skku.edu}

\abstract{
In this study, as an extension of our previous work, we estimate the sensitivity of the Search for Hidden Particles (SHiP) experiment to the 3+1 model using the charged-current deep inelastic scattering event spectrum. 
We employ the Feldman-Cousins method with a parametric bootstrap to account for nuisance parameters and systematic uncertainties.
In the previous study, we proposed a dual baseline approach by suggesting Far SND (FSND) at 120 m with Near SND (NSND) at 27 m. We employ the same approach in this study. 
The NSND-only configuration can probe mixing parameters of $|U_{\alpha4}|^2 \gtrsim 0.1$ near $\Delta m_{41}^2 \sim 10^3\,\mathrm{eV}^2$, with a reduction of normalized systematic uncertainties from 20\% to 10\% improving sensitivity by roughly a factor of two. 
Moreover, the inclusion of FSND significantly enhances the sensitivity by a factor of 2 to 10 depending on the flavor and the systematic uncertainty. 
In two-flavor mixing scenarios, a cancellation between neutrino appearance and disappearance generates kinks in the sensitivity curves, that are vanished in the dual-baseline approach.}

\begin{document}
\maketitle
\flushbottom

%%%%%%%%%%%%%%%%%%%%%%%%%%%%%%%%%%
\section{Introduction}
\label{Intro}
%%%%%%%%%%%%%%%%%%%%%%%%%%%%%%%%%%

Neutrino oscillation experiments have shown that neutrinos have masses and mixing among three active flavors~\cite{ParticleDataGroup:2024cfk}, which we call the three-flavor model.
Long baseline experiments that measure solar neutrinos~\cite{Kamiokande-II:1991pyu,GALLEX:1996xbd,Super-Kamiokande:2001ljr,SNO:2002tuh,SAGE:2009eeu,Borexino:2013zhu}, atmospheric neutrinos~\cite{Super-Kamiokande:1998kpq,ANTARES:2012tem,IceCube:2013pav}, and neutrinos that are produced by colliders~\cite{MINOS:2008kxu,T2K:2013bzi,NOvA:2016vij,OPERA:2018nar} probe the three-flavor model from various perspectives.
However, a recent result from IceCube~\cite{IceCube:2024pky} or anomalies observed in LSND~\cite{LSND:2001aii}, MiniBooNE~\cite{MiniBooNE:2020pnu}, reactor neutrino experiments~\cite{Mention:2011rk} and gallium experiments~\cite{Elliott:2023cvh} could suggest the existence of an additional neutrino flavor state that is \textit{sterile}, meaning it does not participate in weak interactions.

Extensions of the Standard Model of particle physics that include sterile neutrinos may address open questions such as the origin of the smallness of neutrino masses~\cite{Minkowski:1977sc,Ramond:1979py,Gell-Mann:1979vob,Yanagida:1979as,Mohapatra:1979ia}, the nature of dark matter~\cite{Boyarsky:2018tvu}, and the baryon asymmetry of the Universe~\cite{Asaka:2005pn}.
The simplest extension of the Standard Model with additional sterile flavor and mass eigenstate is called the ``3+1" model~\cite{Acero:2022wqg}, and several experiments have constrained the model.
Long-baseline experiments such as Super-Kamiokande~\cite{Super-Kamiokande:2014ndf}, NOvA~\cite{NOvA:2017geg}, MINOS~\cite{MINOS:2017cae}, T2K~\cite{T2K:2019efw}, OPERA~\cite{OPERA:2019kzo}, and IceCube-DeepCore~\cite{IceCube:2024dlz} have excluded $|U_{\alpha4}|^2 \gtrsim 0.1$ regions with $\Delta m_{41}^2 \gtrsim 0.1\,\mathrm{eV}^2$. 
Moreover, reactor neutrino experiments~\cite{DANSS:2018fnn,NEOS:2016wee,HOUMMADA1995449,PROSPECT:2018dtt,STEREO:2018rfh,RENO:2020hva,RENO:2020uip,DayaBay:2024nip} constrain $|U_{e4}|^2 \gtrsim 0.1$ for $\Delta m_{41}^2 \lesssim 10\,\mathrm{eV}^2$, while tritium beta decay experiments~\cite{Kraus:2012he,Belesev:2012hx,Belesev:2013cba,Abdurashitov:2017kka,KATRIN:2022ith,KATRIN:2022spi} constrain $|U_{e4}|^2 \gtrsim 10^{-2}$ for $\Delta m_{41}^2 \gtrsim 10^2\,\mathrm{eV}^2$.

Short-baseline experiments, particularly those using fixed-target neutrino sources, have also played a crucial role in probing sterile neutrinos~\cite{DYDAK1984281,PhysRevLett.52.1384,CCFRNuTeV:1998gjj,NOMAD:2001xxt,NuTeV:2002daf,NOvA:2017geg,KARMEN:2002zcm,NOMAD:2003mqg,CHORUS:2007wlo,MiniBooNE:2012meu,MicroBooNE:2022sdp}.
Notably, SPS-based experiments, such as CDHS~\cite{DYDAK1984281}, NOMAD~\cite{NOMAD:2001xxt,NOMAD:2003mqg} and CHORUS~\cite{CHORUS:2007wlo}, as well as Tevatron-based experiments such as CCFR~\cite{PhysRevLett.52.1384,CCFRNuTeV:1998gjj} and NuTeV~\cite{NuTeV:2002daf} provided constraints on the $\Delta m_{41}^2\gtrsim100\,\mathrm{eV}^2$ region. 
Their high energy scales generate a detectable flux of tau neutrinos, which is why only a few experiments at the SPS or Tevatron have probed tau neutrino appearance on a short baseline.
For higher $\Delta m_{41}^2$ regions, forward neutrino experiments at the LHC can probe the $\Delta m_{41}^2 \gtrsim 10^3\,\mathrm{eV}^2$ region~\cite{Bai:2020ukz,FASER:2019dxq} which is our interest in later contents.

The Search for Hidden Particles (SHiP) experiment has been proposed to search for feebly interacting GeV-scale particles and to perform measurements in neutrino physics at the CERN SPS.~\cite{SHiP:2015vad,Alekhin:2015byh,Ahdida:2654870,Albanese:2878604}
Like other SPS-based experiments, SHiP could serve as an additional probe of the 3+1 model as we discussed in our previous study~\cite{Choi:2024ips}.
SHiP will collide $400\,\mathrm{GeV}$ SPS protons on a heavy target to produce an intense neutrino beam containing all flavors, including a significant flux of $\nu_\tau$. 
Then the Scattering and Neutrino Detector (SND) at SHiP, which utilizes an emulsion cloud chamber and a subsequent muon spectrometer for both tracking and calorimetry, is designed to identify charged-current interactions and distinguish neutrino flavors with high spatial resolution.

In this study, as an extension of our previous study~\cite{Choi:2024ips}, we find the sensitivities of the 3+1 model using neutrino Charged Current Deep Inelastic Scattering (CC DIS) event spectrum and compare the sensitivities with current constraints from past experiments.
We also investigate the impact of a secondary SND detector at a $120\,\mathrm{m}$ baseline that is proposed on our previous study~\cite{Choi:2024ips}, which would enable us to cross-check systematic uncertainties and the oscillation pattern, therefore enhance sensitivities to the 3+1 model. 

The paper is organized as follows: In Section~\ref{sec:3+1}, we review the 3+1 sterile neutrino model in the context of SHiP’s experimental setup. Section~\ref{sec:method} details the statistical framework and assumptions used to derive sensitivity estimates. 
Section~\ref{sec:sensitivity} presents our results and compares them with existing constraints. 
Finally, Section~\ref{sec:conclusion} summarizes our findings and discusses future prospects.

%%%%%%%%%%%%%%%%%%%%%%%%%%%%%%%%%%%%%%%%%%%%%%%%%%%%
\section{The 3+1 model on the SHiP experiment}
\label{sec:3+1}
%%%%%%%%%%%%%%%%%%%%%%%%%%%%%%%%%%%%%%%%%%%%%%%%%%%%

In this section, we review the phenomenological aspects of sterile neutrinos within the context of SHiP.

The weak eigenstates of neutrinos $\ket{\nu_\alpha}$, including sterile flavors in general, can be transformed into mass eigenstates $\ket{\nu_i}$ using the mixing matrix $\B{U}$,
\begin{equation}
\ket{\nu_\alpha}=\sum_{i} U^{\ast}_{\alpha i}\ket{\nu_i}.
\end{equation}
The Standard Model of particle physics includes three active neutrino flavors, while various extensions introduce additional sterile flavors.
In the 3+1 model, the mixing matrix becomes $4\times 4$ matrix with an additional single sterile flavor:
\begin{equation}
\B{U}=\mathcal{U}_{34}\mathcal{V}_{24}\mathcal{V}_{14}\mathcal{U}_{23}\mathcal{V}_{13}\mathcal{U}_{12}=
\begin{pmatrix}
U_{e 1} & U_{e 2} & U_{e 3} & U_{e 4}\\
U_{\mu 1} & U_{\mu 2} & U_{\mu 3} & U_{\mu 4}\\
U_{\tau 1} & U_{\tau 2} & U_{\tau 3} & U_{\tau 4}\\
U_{s 1} & U_{s 2} & U_{s 3} & U_{s 4}
\end{pmatrix},
\end{equation}
where, for example, rotation matrices $\mathcal{U}_{12}$ and $\mathcal{V}_{13}$ are defined as~\cite{HARARI1986123}
\begin{equation}
\mathcal{U}_{12}=\begin{pmatrix}
\cos\theta_{12} & \sin\theta_{12} & 0 & 0\\
-\sin\theta_{12} & \cos\theta_{12} & 0 & 0\\
0 & 0 & 1 & 0\\
0 & 0 & 0 & 1
\end{pmatrix},
\mathcal{V}_{13}=\begin{pmatrix}
\cos\theta_{13} & 0 & \sin\theta_{13}e^{-i\delta_{13}} & 0\\
0 & 1 & 0 & 0\\
-\sin\theta_{13}e^{i\delta_{13}} & 0 & \cos\theta_{13} & 0\\
0 & 0 & 0 & 1
\end{pmatrix}.
\end{equation}
Such parameterization is necessary when neutrino oscillations from the three-flavor model are effective, but it can be simplified for short baseline experiments.

Under the relativistic approximation, if the mass of the extra mass eigenstate is sufficiently large and the baseline is short enough to neglect lighter mass eigenstates, the transition probability from an active flavor $\alpha$ to another active flavor $\beta$ can be approximated as
\begin{equation}\label{eq:short}
P_{\alpha \beta} \approx \left|\delta_{\alpha \beta} -\left(1-e^{-i\frac{\Delta m^2_{41} L}{2 E_\nu}}\right)U_{\alpha 4} U_{\beta 4}^\ast\right|^2,
\end{equation}
where $L$ is the baseline, and $E_\nu$ is the neutrino energy.
Introducing the notation $\sin^22\theta_{\alpha\beta} \equiv 4|U_{\alpha4}|^2(1-|U_{\alpha4}|^2)$ for $\alpha = \beta$ and $\sin^22\theta_{\alpha\beta} \equiv 4|U_{\alpha4}|^2|U_{\beta4}|^2$ for $\alpha\neq\beta$, Eq.~(\ref{eq:short}) can be simplified into
\begin{equation}\label{eq:short_simple}
P_{\alpha\beta} =
\begin{cases}
1-\sin^22\theta_{\alpha\beta} \sin^2\left(\frac{\Delta m^2_{41} L}{4 E_\nu}\right) & \text{if } \alpha=\beta\\
\sin^22\theta_{\alpha\beta} \sin^2\left(\frac{\Delta m^2_{41} L}{4 E_\nu}\right) & \text{if } \alpha\neq\beta.
\end{cases}
\end{equation}
Thus, under the short baseline approximation, the 3+1 model can be parameterized by the mass-squared difference $\Delta m_{41}^2$ and the mixing parameters $|U_{\alpha4}|^2$ while every CP violating phase can be ignored.

\begin{figure}
\centering
\includegraphics[width = 0.7\textwidth]{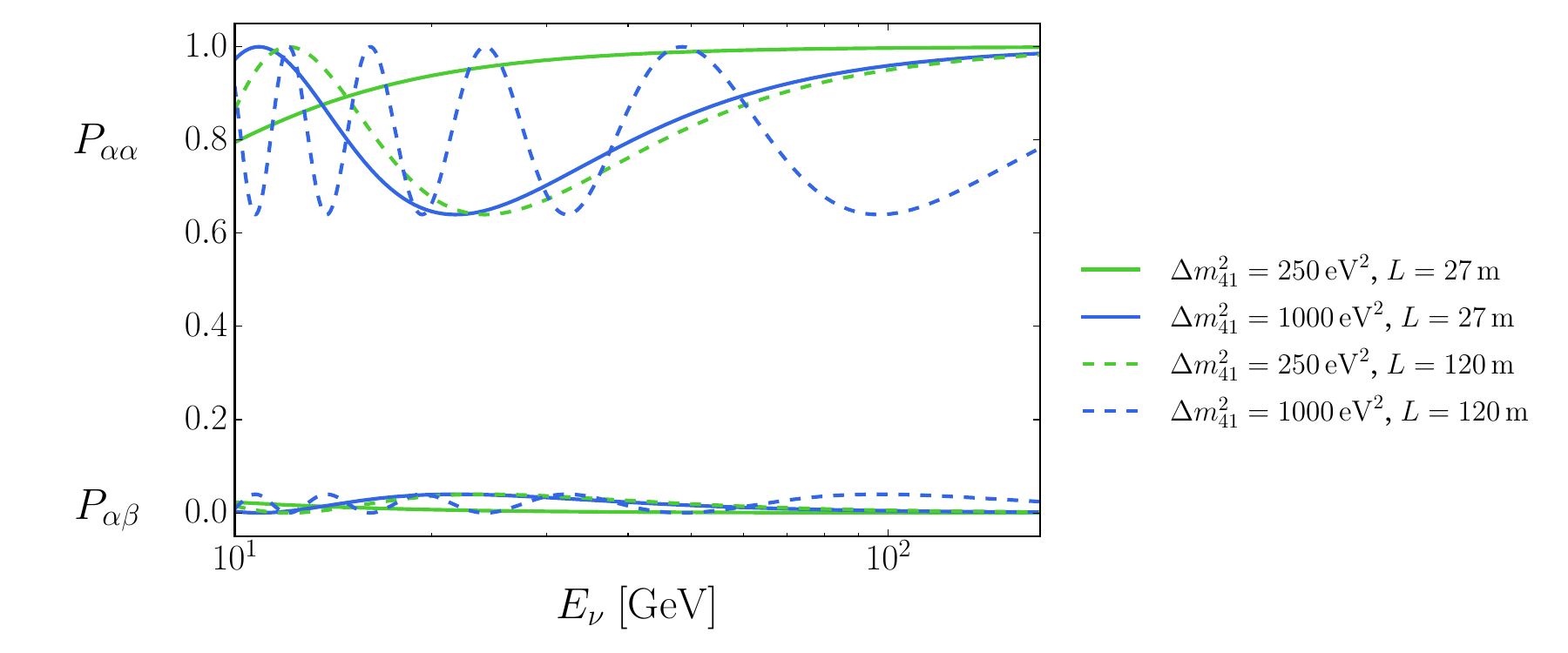}
\caption{Appearance (bottom) and disappearance (top) probabilities at $27\,\mathrm{m}$ (solid) and $120\,\mathrm{m}$ (dashed) baselines for mass-squared differences $\Delta m_{41}^2 = 250\,\mathrm{eV}^2$ (green) and $\Delta m_{41}^2 = 10^3\,\mathrm{eV}^2$ (blue), with mixing parameters $|U_{\alpha4}|^2 = |U_{\beta4}|^2 = 0.1$.}
\label{fig:osc}
\end{figure}

The maximum sensitivity is achieved near $\Delta m_{41}^2 \sim 2\pi \braket{E_\nu}/L$, where $\braket{E_\nu}$ denotes the average neutrino energy produced by the proton target. 
In this paper, following Ref.~\cite{Choi:2024ips}, we refer to the SND with a $27\,\mathrm{m}$ baseline as Near SND (NSND) and the SND with a $120\,\mathrm{m}$ baseline as Far SND (FSND).
Assuming an average neutrino energy of $30\,\mathrm{GeV}$, NSND is most effective at $\Delta m_{41}^2\sim10^3\,\mathrm{eV}^2$, while FSND is most effective at $\Delta m_{41}^2\sim250\,\mathrm{eV}^2$.
Fig.~\ref{fig:osc} depicts probabilities of appearance (bottom) and disappearance (top) at $27\,\mathrm{m}$ (solid) and $120\,\mathrm{m}$ (dashed) baselines for mass-squared differences $\Delta m_{41}^2 = 250\,\mathrm{eV}^2$ (green) and $\Delta m_{41}^2 = 10^3\,\mathrm{eV}^2$ (blue), with mixing parameters $|U_{\alpha4}|^2 = |U_{\beta4}|^2 = 0.1$.
Note that the total oscillation probability sums to unity when the active–sterile transition probability is included.

%%%%%%%%%%%%%%%%%%%%%%%%%%%%%%%%%%%%%%%%%%%%%%%%%%%%
\section{Method on sensitivity estimation}
\label{sec:method}
%%%%%%%%%%%%%%%%%%%%%%%%%%%%%%%%%%%%%%%%%%%%%%%%%%%%
In this section, we present the method for estimating the sensitivity of SHiP under the 3+1 model.
This sensitivity is determined by identifying the confidence region for the data, assuming that both SHiP and auxiliary measurements observed the most probable outcome under the three-flavor model.
To determine the confidence region, we apply the Feldman--Cousins (FC) method~\cite{Feldman:1997qc} with a parametric bootstrap~\cite{Demortier2012}, which constructs a confidence region for a given sample distribution. 
In real analyses, repeated sampling for SHiP is performed using Monte Carlo (MC) simulation tools such as FairSHiP~\cite{Albanese:2878604}, which is beyond the main focus of our study.
Instead, a crude estimation of the probability distribution for the SHiP sample is performed, and a confidence region is determined accordingly.

This section begins with a review of the FC method with a parametric bootstrap to find a confidence region with nuisance parameters. 
Next, we estimate the probability distribution of samples and point estimates of nuisance parameters for SHiP.

\subsection{Statistical method}

In this subsection, we introduce the statistical methods that we employed to estimate the sensitivities.
The confidence region with significance level $\alpha \equiv 1-\mathrm{CL}$ using the FC method is defined as follows:
\begin{equation}
\mathrm{CR}(\B{D}, \B{\varphi}, \alpha) = \{ \B{\theta} | \mathfrak{p}(\B{\theta}; \B{D}, \B{\varphi}) > \alpha \},
\end{equation}
where $\B{D}$ represents the data from a SND, $\B{\varphi}$ is the point estimate of the nuisance parameters $\B{\phi}$ obtained from auxiliary measurements, and $\mathfrak{p}(\B{\theta}; \B{D}, \B{\varphi})$ denotes the p-value for the model parameters $\B{\theta}$ given $\B{D}$ and $\B{\varphi}$. 
Ideally, repeated sampling for auxiliary measurements should be performed to follow the FC method rigorously, although this is often not the case in practice. 
Therefore, we utilize a parametric bootstrap~\cite{Demortier2012} for the nuisance parameters, treating $\B{\varphi}$ as a substitute for the data of auxiliary measurements.

The p-value is defined as
\begin{equation}
1 - \mathfrak{p}(\B{\theta}; \B{D}, \B{\varphi}) = \int_0^{\lambda(\B{D}, \B{\varphi}, \B{\theta})} \rho(\lambda' | \B{\theta}, \B{\phi}) d\lambda',
\end{equation}
where $\rho(\lambda' | \B{\theta}, \B{\phi})$ is the probability density function of the test statistic $\lambda'$ for given $\B{\theta}$ and $\B{\phi}$, which is determined through repeated sampling of $\B{D}$ and $\B{\varphi}$.
If we denote the samples of $\B{D}$ and $\B{\varphi}$ as $\B O$ and $\B\varphi'$, the probability of the test statistic is equal to
\begin{equation}
    \rho(\lambda' | \B{\theta}, \B{\phi}) = \int \delta[\lambda' - \lambda(\B O,\B\varphi',\B\theta)] \rho(\B O,\B\varphi'|\B\theta, \B\phi) d\B O\, d\B\varphi',
\end{equation}
where $\delta(x-y)$ is a Dirac-delta distribution and $\rho(\B O,\B\varphi'|\B\theta, \B\phi)$ is the sample distribution of SHiP and its auxiliary measurements.

The sample distribution in this paper is written as:
\begin{equation}
    \rho(\B O,\B\varphi'|\B\theta, \B\phi) = \prod_{\alpha,i} \mathrm{Pois}(O_{\alpha,i}|\mu_{\alpha,i}) \times \rho(\B\varphi'|\B\phi),
\end{equation}
where $\mu_{\alpha,i}$ is the mean number of events at the $i$-th bin of $\alpha$ flavor CC DIS channel which is a function of $\B\theta$ and $\B\phi$, 
$O_{\alpha,i}\in\B O$ is the number of events of the sample at the $i$-th bin of $\alpha$ flavor CC DIS channel, 
$\mathrm{Pois}(x|\mu)$ is a Poisson distribution of $x$ for mean $\mu$,
and $\rho(\B\varphi'|\B\phi)$ is the probability density function of point estimates of $\B\phi$ which determines the systematic uncertainty of SHiP.
For the test statistic, the profile likelihood ratio~\cite{stuart1999kendalls,Cousins:2018tiz} is chosen which is defined as follows:
\begin{equation}
\lambda(\B{O}, \B{\varphi}, \B{\theta}) \equiv -2 \log \frac{\max_{\B{\phi}} \rho(\B{O}, \B{\varphi} | \B{\theta}, \B{\phi})}{\max_{\B{\theta}, \B{\phi}} \rho(\B{O}, \B{\varphi} | \B{\theta}, \B{\phi})}.
\end{equation}

However, to perform repeated sampling, one must choose specific values of $\B\theta$ and $\B\phi$ while the definition of the p-value does not specify $\B\phi$.
Several variations of the FC method, such as the Berger--Boos method~\cite{Berger01091994} or the Highland--Cousins method~\cite{COUSINS1992331} are proposed to solve the issue.
In this paper, we use the profiled FC method which selects the value $\B\phi$ that maximize $\rho(\B{D}, \B{\varphi} | \B{\theta}, \B{\phi})$ during repeated sampling.~\cite{Cranmer:2014lly, NOvA:2022wnj}

\subsection{Estimation of SHiP data}

In this subsection, we estimate the number of CC DIS events at SHiP and the probability distribution of SHiP samples.
If event classification were perfect, $\mu_{\alpha,i}$ would equal to the mean number of actual CC DIS events with the corresponding flavor.
However, due to imperfections in event classification, some CC DIS events could be excluded by the criteria of the corresponding channel or even classified as a different flavor, and events from other sources could be classified as CC DIS events.
Therefore, only a portion of actual CC DIS events with corresponding flavor contributes as a signal, while contributions from other sources constitute the background, which can be written as follows: 
\begin{equation}
    \mu_{\alpha,i} \equiv s_{\alpha,i} + b_{\alpha,i},
\end{equation}
where $s_{\alpha,i}$ and $b_{\alpha,i}$ are the signal and the background at the $i$-th bin of $\alpha$ flavor CC DIS channel. 
Here, we estimate the signal $s_{\alpha,i}$ using the formula below:
\begin{equation}
\label{eq:signal}
    s_{\alpha,i} = \sum_{\beta}^{e,\mu,\tau}(1+\phi_\beta + \phi_{\beta, i})\int_{E_{\mathrm{start},i}}^{E_{\mathrm{end},i}}dE_\nu\int_{L_0}^{L_0+L_\mathrm{d}}dL\int_{0}^{\infty}dE_\mathrm{rec}\varepsilon_\alpha P_{\beta\alpha}\frac{d^2N_{\beta}}{dE_\nu dL}\frac{\sigma_{\nu_\alpha A}}{\sigma_{\nu_\beta A}} \rho(E_\mathrm{rec}|E_\nu),
\end{equation}
where parameters are defined as:
\begin{itemize}
    \item $\phi_\beta\in\B\phi$ and  $\phi_{\beta, i} \in \B\phi$ are overall and shape nuisance parameters for the neutrino flux with flavor $\beta$,
    \item $N_\beta$ is the number of neutrinos with flavor $\beta$ that pass through a SND under the three-flavor hypothesis,
    \item each $\varepsilon_\alpha$ and $\sigma_{\nu_\alpha A}$ are a detection efficiency and the scattering cross section of neutrino CC DIS events with flavor $\alpha$,
    \item $L_0$ is a distance between the proton target and the detector,
    \item $L_\mathrm{d}$ is a length of the detector,
    \item $E_\mathrm{rec}$ is a reconstructed energy of the neutrino that is estimated by the data from a SND, 
    \item $[E_{\mathrm{start},i},E_{\mathrm{end},i}]$ is the range of reconstructed energy for $i$-th bin,
    \item and $\rho(E_\mathrm{rec}|E_\nu)$ is a probability density of $E_\mathrm{rec}$ for the given $E_\nu$, which is chosen as a Gaussian distribution with the standard deviation $\sigma=0.2 E_\nu$ following the analysis of Ref.~\cite{Buonaura:2268663}.
\end{itemize}

In this study, we use a total of 7 energy bins according to the reconstructed energy.
Each energy bin starts at the endpoint of the previous energy bin and ends at 1.5 times its starting energy, with the first energy bin beginning at $10\,\mathrm{GeV}$.
Additionally, to ensure that $s_{\alpha,i}$ remains positive, the domain of each nuisance parameter is chosen as $(-0.5,\infty)$, and the probability density function of the point estimate of nuisance parameters is chosen as
\begin{equation}\label{eq:nuisance}
     \rho(\B\varphi|\B\phi) = \prod_{j}\mathcal{N}_{(-0.5,\infty)}(\varphi_{j}|\phi_j,\sigma_{\mathrm{sys},j}),
\end{equation}
where $\mathcal{N}_{(a,b)}(x|\mu,\sigma)$ is a truncated normal distribution with the mean value $\mu$ and the standard deviation $\sigma$ on the domain of $(a,b)$, and $\varphi_{j}$ and $\phi_{j}$ are $j$-th components of $\B{\varphi}$ and $\B{\phi}$.
In general cases, covariance terms between different nuisance parameters are included in systematic uncertainties. 
In this study, to simplify the situation, we ignore any covariance terms inside $\rho(\B\varphi|\B\phi)$ and $\sigma_{\mathrm{sys},j}$ are chosen to be equal to $\sigma_\mathrm{norm}$, as the biggest contribution of the uncertainty comes from the neutrino flux.
The current uncertainty of the tau neutrino flux that is generated from the proton target is $\sim40\%$~\cite{Bai:2018xum}, while DsTau could improve the uncertainty around $10\%$.~\cite{DsTau:2019wjb}

The estimated number of neutrino CC DIS events is given by repeated sampling of SHiP, while background and detection efficiencies are estimated by testing particle identification methods on MC samples from repeated sampling. 
In this study, instead of rigorous repetitive sampling of SHiP, we import the number of CC DIS events from the recent report of SHiP~\cite{Albanese:2878604} assuming that the differential number of events is constant with respect to the distance.
Also, to estimate the background and detection efficiencies of CC DIS events, we use results from OPERA~\cite{OPERA:2019kzo} for electron and muon neutrino channels.
For the tau neutrino channel, on the other hand, we use the analysis of Ref.~\cite{Iuliano:2776128} with the criterion $\phi_{\tau-h} \geq 2.8\,\mathrm{rad}$ where $\phi_{\tau-h}$ is the parameter between the hadronic jet and the tau lepton track on the plane that is perpendicular with neutrino vertex. 
The detection efficiencies are therefore chosen as follows:
\begin{equation}
\begin{split}
\varepsilon_{e} &= 30\%\\
\varepsilon_{\mu} &= 40\%\\
\varepsilon_{\tau} &= 10\%,
\end{split}
\end{equation}
where we assume no dependence on the incoming neutrino energy.

Regarding background contributions, we assume electron and muon neutrino CC DIS channels contain negligible background as OPERA achieved purities of $\sim96\%$ and $99.5\%$~\cite{OPERA:2019kzo}, respectively.
For tau neutrino channel, the source of background is dominated by muon neutrino CC events that generate charm mesons while outgoing muons are unidentified.
In this study, we choose a background of the tau neutrino CC DIS channel with an energy-independent signal-to-background ratio $R_\mathrm{s/b} = 2$ for three-flavor model.
Therefore, $b_{\tau,i}$ can be written as
\begin{equation}
\label{eq:bkg}
    b_{\tau,i} = (1+\phi_\mu + \phi_{\mu, i})\int_{E_{\mathrm{start},i}}^{E_{\mathrm{end},i}}dE_\nu\int_{L_0}^{L_0+L_\mathrm{d}}dL\int_{0}^{\infty}dE_\mathrm{rec}R_\mathrm{s/b}^{-1}\varepsilon_\tau P_{\mu\mu}\frac{d^2N_{\tau}}{dE_\nu dL} \rho(E_\mathrm{rec}|E_\nu),
\end{equation}
where we ignore any background contribution from muon neutrino appearance as they are negligible compared with the signal contribution from tau neutrino appearance. 
Note that detection efficiencies and purities of each channels can be enhanced by recent advancements in computing power and machine learning~\cite{Feickert:2021ajf}, which is not considered in this paper.

To calculate the contribution of neutrino appearance, we consider the ratio of neutrino CC DIS cross section between two flavors, following Ref.~\cite{Bai:2018xum}.
The ratio of neutrino CC DIS cross section remains the same for anti-neutrinos, as the ratio depends on the mass of the lepton and the energy of the incoming neutrino.
We also ignore the difference between $\sigma_{\nu_e A}$ and $\sigma_{\nu_\mu A}$ since the masses of the electron and muon are negligible in our energy regime.

\begin{figure}
    \centering
    \includegraphics[width = .99\textwidth]{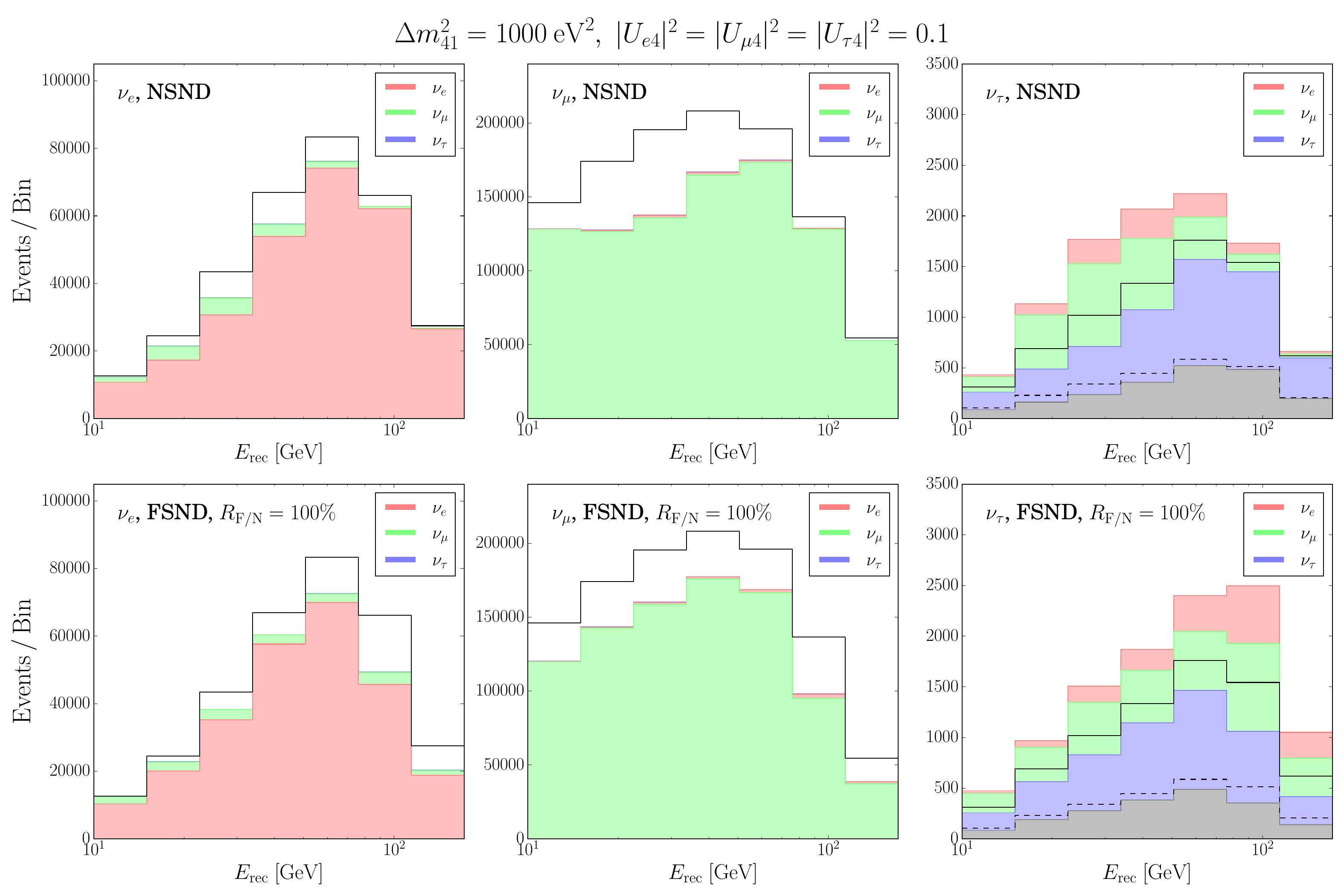}
    \caption{Histograms showing the mean number of CC DIS events per energy bin for each electron (left), muon (center) and tau (right) flavor channel in NSND (up) and FSND with $R_\mathrm{F/N}=100\%$ (down), assuming null nuisance parameters. 
    Colored regions correspond to the contributions of $e$ (red), $\mu$ (green) and $\tau$ (blue) flavor neutrino flux under the 3+1 model with $\Delta m_{41}^2=1000\,\mathrm{eV}^2$ and $|U_{e4}|^2=|U_{\mu4}|^2=|U_{\tau4}|^2=0.1$, following Eq.~(\ref{eq:signal}).
    Gray regions in the $\tau$ flavor channels indicate background contributions under the same $\Delta m_{41}^2$ and mixing parameters, following Eq.~(\ref{eq:bkg}).
    For comparison, solid and dashed black lines represent the corresponding mean number of events and background under the standard three-flavor model.
    While for electron and muon flavor channels, backgrounds are assumed as zero as they were negligible in OPERA.
    }
    \label{fig:binned}
\end{figure}

Finally, gathering more data from FSND would increase the sensitivity of the 3+1 model due to its longer baseline compared to NSND.
For simplicity, we assume that the signal and the background of neutrino CC DIS events at FSND follow the same formula as those at NSND, using the same nuisance parameters, while the signal and the background at FSND are multiplied by the overall ratio $R_\mathrm{F/N}$ to reflect scaling due to the detector configuration.
Also, we choose $L_0=27\,\mathrm{m}$ and $L_\mathrm{d} = 2.6\,\mathrm{m}$ for NSND, and $L_0=120\,\mathrm{m}$ and $L_\mathrm{d} = 2.6\,\mathrm{m}$ are chosen for FSND.
In reality, uncertainties due to the properties of detectors reduce the correlation between NSND and FSND, which may contribute to a reduction in the sensitivity compared to our estimation.

Fig.~\ref{fig:binned} shows histograms of the mean number of CC DIS events per energy bin for each electron (left), muon (center) and tau (right) flavor channel in NSND (up) and FSND with $R_\mathrm{F/N}=100\%$ (down). 
Colored regions show contributions of $e$ (red), $\mu$ (green) and $\tau$ (blue) flavor neutrino flux on CC DIS event channels under the 3+1 model with $\Delta m_{41}^2=1000\,\mathrm{eV}^2$ and $|U_{e4}|^2=|U_{\mu4}|^2=|U_{\tau4}|^2=0.1$.
Grey regions in $\tau$ flavor channels indicate background contributions under the same $\Delta m_{41}^2$ and mixing parameters, assuming that all background consists of muon neutrinos.
Mean numbers of event are derived by Eq.~(\ref{eq:signal}) and Eq.~(\ref{eq:bkg}), while every nuisance parameter is assumed to be zero.
For comparison, we also draw mean numbers of event and background for the three-flavor model with null nuisance parameters as black solid lines and black dashed lines, which will be used as a data to find a sensitivity. 
Notably, neutrino disappearance on tau flavor is compensated by the appearance from other flavors, which indicates that in specific mixing parameters, there could be a cancellation between the appearance and the disappearance if sterile neutrinos are mixed with two or more active flavors.
We will examine the impact of this cancellation effect in detail in the next section.

%%%%%%%%%%%%%%%%%%%%%%%%%%%%%%%%%%%%%%%%%%%%%%%%%%%%
\section{Estimated Sensitivities of SHiP on the 3+1 model }
\label{sec:sensitivity}
%%%%%%%%%%%%%%%%%%%%%%%%%%%%%%%%%%%%%%%%%%%%%%%%%%%%

In this section, we present the estimated sensitivities of SHiP to the 3+1 model under various combinations of the normalized systematic uncertainty and the FSND contribution. 
To evaluate the sensitivity, we compute 90\% CL confidence regions in the parameter spaces, assuming that the observed number of CC DIS events coincides with the mean expectation under the three-flavor model with zero nuisance parameters. 
First, we consider scenarios in which only a single flavor is mixed with sterile neutrinos, so that only disappearance can be observed. 
In these cases, the sensitivities are presented on the $(\Delta m_{41}^2, |U_{\alpha4}|^2)$ plane for each active flavor $\alpha$. 
Next, we consider scenarios in which two flavors are mixed with sterile neutrinos, and the corresponding sensitivities are shown on the $(|U_{\alpha4}|^2, |U_{\beta4}|^2)$ plane at fixed values of $\Delta m_{41}^2$ with reference values of $50\,\mathrm{eV}^2$, $250\,\mathrm{eV}^2$, $1000\,\mathrm{eV}^2$, and $5000\,\mathrm{eV}^2$.
Note that each $250\,\mathrm{eV}^2$ and $1000\,\mathrm{eV}^2$ are optimal choice of $\Delta m_{41}^2$ for NSND and FSND, while cases with $50\,\mathrm{eV}^2$ and $5000\,\mathrm{eV}^2$ show how the contours evolve away from optimal points.

Every figure in this section is drawn following rules below. 
First, a Fisher $z$-transformation is applied to the p-value to reduce interpolation errors. 
Different settings for the normalized systematic uncertainty are distinguished by line styles: cases with $\sigma_\mathrm{norm}=10\%$ are depicted with dashed lines, while those with $\sigma_\mathrm{norm}=20\%$ are shown with solid lines. 
For the FSND contributions, NSND-only scenarios (i.e., $R_\mathrm{F/N}=0\%$, corresponding to the current design of SHiP) are represented by blue lines, cases with $R_\mathrm{F/N}=10\%$ are drawn in purple, and those with $R_\mathrm{F/N}=100\%$ are drawn in red. 
Also, for comparison with past experiments, we include current constraints from other experiments that did not reject the three-flavor hypothesis with enough margin.
Constraints on the same parameter space are shown as green regions, or as light green hatched regions if constraints are given with additional assumptions.
Because confidence regions in different parameter spaces cannot be directly compared, we show constraints from different parameter spaces as grey regions, where we include such regions only if the limit holds regardless of the values of the other 3+1 model parameters.
Finally, every constraint from past experiments have $90\%$ CL, except tritium beta decay experiments which have $95\%$ CL.

\subsection{Single flavor mixing with sterile neutrino}

\begin{figure}
    \centering
    \includegraphics[width=.99\textwidth]{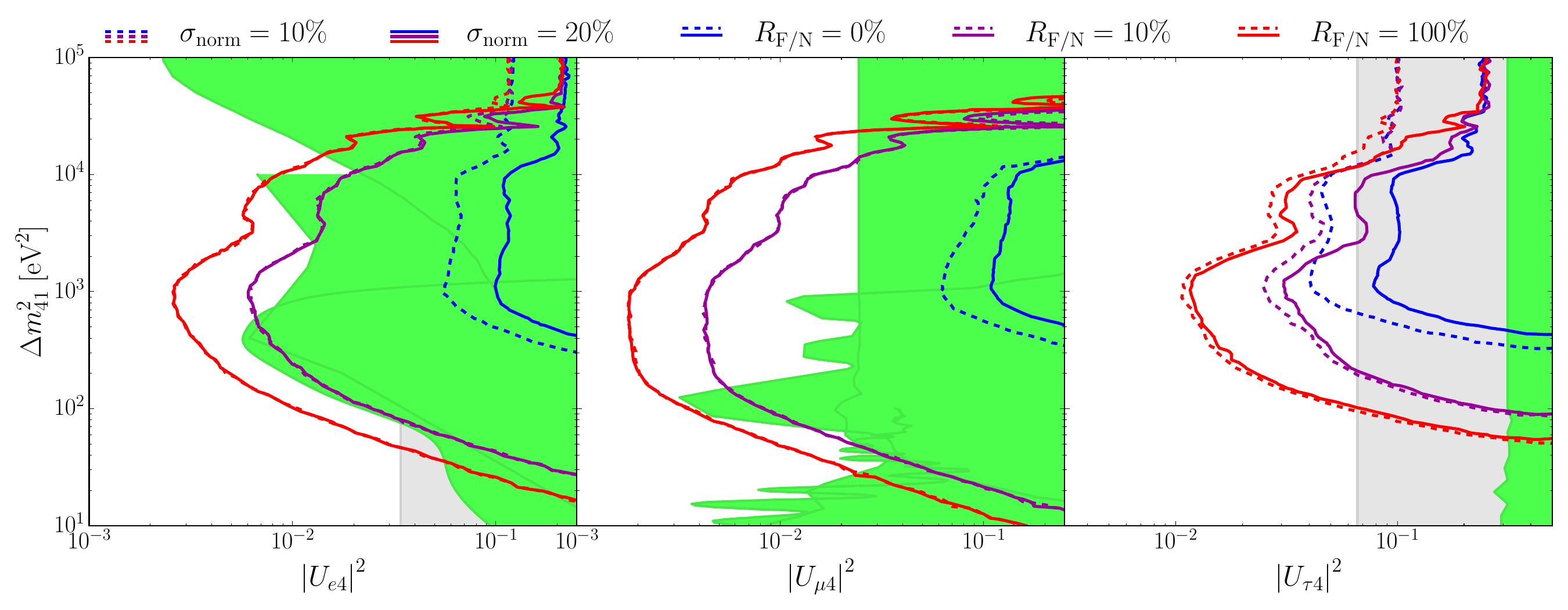}
    \caption{Expected sensitivities (90\% CL) of SHiP after a 5-year operation in the $(|U_{\alpha 4}|^2,\,\Delta m^2_{41})$ plane, assuming zero mixing with flavors other than $\alpha$. 
    Sensitivities are shown for $\sigma_\mathrm{norm}=10\%$ (dashed) and $\sigma_\mathrm{norm}=20\%$ (solid), while $R_\mathrm{F/N}$ is varied as $0\%$ (blue, NSND-only), $10\%$ (purple), and $100\%$ (red). 
    Green regions are disfavored by past experiments that are from tritium beta decay experiments~\cite{Belesev:2013cba,KATRIN:2022ith,KATRIN:2022spi} for electron flavor, CCFR~\cite{PhysRevLett.52.1384}, MiniBooNE+SciBooNE~\cite{MiniBooNE:2012meu} and MINOS~\cite{MINOS:2017cae} for muon flavor, and NOvA~\cite{NOvA:2024imi} for tau flavor parameter space. 
    Grey regions indicate constraints from different parameter spaces: the limit on $|U_{e4}|^2$ is derived from Super-Kamiokande~\cite{Super-Kamiokande:2014ndf}, while the limit on $|U_{\tau4}|^2$ is derived from IceCube-DeepCore~\cite{IceCube:2024dlz}.}
    \label{fig:disappearance}
\end{figure}

Fig.~\ref{fig:disappearance} shows the expected sensitivities (90\% CL) of the SHiP experiment in the $(|U_{\alpha 4}|^2,\,\Delta m^2_{41})$ plane after a 5-year operation, assuming zero mixing with flavors other than $\alpha$. 
Sensitivities are shown for $\sigma_\mathrm{norm}=10\%$ (dashed) and $\sigma_\mathrm{norm}=20\%$ (solid), while $R_\mathrm{F/N}$ is varied as $0\%$ (blue, NSND-only), $10\%$ (purple), and $100\%$ (red). 
Green regions are disfavored by past experiments that are from tritium beta decay experiments~\cite{Belesev:2013cba,KATRIN:2022ith,KATRIN:2022spi} for electron flavor, CCFR\cite{PhysRevLett.52.1384}, MiniBooNE+SciBooNE~\cite{MiniBooNE:2012meu} and MINOS~\cite{MINOS:2017cae} for muon flavor, and NOvA~\cite{NOvA:2024imi} for tau flavor parameter space. 
Constraints from MINOS~\cite{MINOS:2017cae} and NOvA~\cite{NOvA:2024imi} are extended to larger $\Delta m_{41}^2$ since they provide constraints at sufficiently large $\Delta m_{41}^2$ where oscillations are averaged on both near and far detectors. 
Grey regions indicate the constraints from different parameter spaces: limit on $|U_{e4}|^2$ is derived from Super-Kamiokande~\cite{Super-Kamiokande:2014ndf} in the $(|U_{e4}|^2,|U_{\mu4}|^2)$ plane assuming $\delta_{24}=0$, while the limit on $|U_{\tau4}|^2$ is derived from IceCube-DeepCore~\cite{IceCube:2024dlz} in the $(|U_{\mu4}|^2,|U_{\tau4}|^2)$ plane.

In the NSND-only case, each flavor probes roughly the $|U_{\alpha4}|^2\gtrsim0.1$ region near $\Delta m_{41}^2\sim10^3\,\mathrm{eV}^2$, and reducing $\sigma_\mathrm{norm}$ from $20\%$ to $10\%$ improves the sensitivity by approximately a factor of 2.
Also, the sensitivities are similar across all flavors in this case because the statistical uncertainty is negligible compared with the systematic uncertainty. 
The sensitivity presented here is stronger than that in our previous study~\cite{Choi:2024ips} since a parametric bootstrap for nuisance parameters was not applied previously.

When including FSND, the sensitivity to neutrino disappearance is unchanged with respect to $\sigma_\mathrm{norm}$ for $\Delta m_{41}^2 \lesssim 10^4\,\mathrm{eV}^2$.
In regions where oscillations are averaged (i.e., $\Delta m_{41}^2 \gtrsim 10^4\,\mathrm{eV}^2$), the sensitivity depends solely on $\sigma_\mathrm{norm}$. 
For scenarios with $R_\mathrm{F/N} = 10\%$, the sensitivities to the electron and muon neutrino mixing parameters are enhanced by roughly a factor of 10 near $\Delta m_{41}^2=10^3\,\mathrm{eV}^2$ compared to NSND-only cases, while for the tau flavor the improvement is by a factor of 2 to 3, depending on $\sigma_\mathrm{norm}$. 
Increasing $R_\mathrm{F/N}$ from $10\%$ to $100\%$ further enhances the sensitivity by approximately a factor of 2.3 in the relevant regions.

Note that sensitivities of cases with FSND are independent of $\sigma_\mathrm{norm}$ if neutrino oscillation is not averaged, which indicates that the cross check with multiple baselines could distinguish the oscillation effect from the fluctuations due to systematic uncertainty.
If the neutrino oscillation is averaged in both near and far detectors, however, sensitivity is only dependent on $\sigma_\mathrm{norm}$.
This is because neutrino oscillations and systematic uncertainties cannot be measured independently in these cases so that there is no more merit on FSND.
Also, the Wilks' theorem with two degrees of freedom is satisfied on the non-averaged regions of electron and tau flavor mixing cases, while for muon flavor mixing cases, the number of effective degrees of freedom is approximately $4$.

Comparing with other constraints, the NSND-only sensitivities are generally not competitive, except for the tau flavor at $\sigma_\mathrm{norm}=10\%$, where the sensitivity exceeds the limit from IceCube-DeepCore~\cite{IceCube:2024dlz}. 
In contrast, scenarios with $R_\mathrm{F/N}\geq10\%$ can probe regions beyond the reach of past experiments. 
Near $\Delta m_{41}^2\sim10^3\,\mathrm{eV}^2$, the sensitivities with $R_\mathrm{F/N}=100\%$ are up to approximately 10 times stronger for the muon flavor and 7 times stronger for the tau flavor than the existing constraints, while for the electron flavor most of the parameter space accessible with $R_\mathrm{F/N}=10\%$ is already constrained by beta decay experiments.
Even with $R_\mathrm{F/N}=100\%$, the sensitivity of SHiP is only about 3 times stronger than the constraints from tritium beta decay experiments.

\subsection{Two-flavor mixing with sterile neutrino}
\label{subsec:two}

\begin{figure}
    \centering
    \includegraphics[width=.99\textwidth]{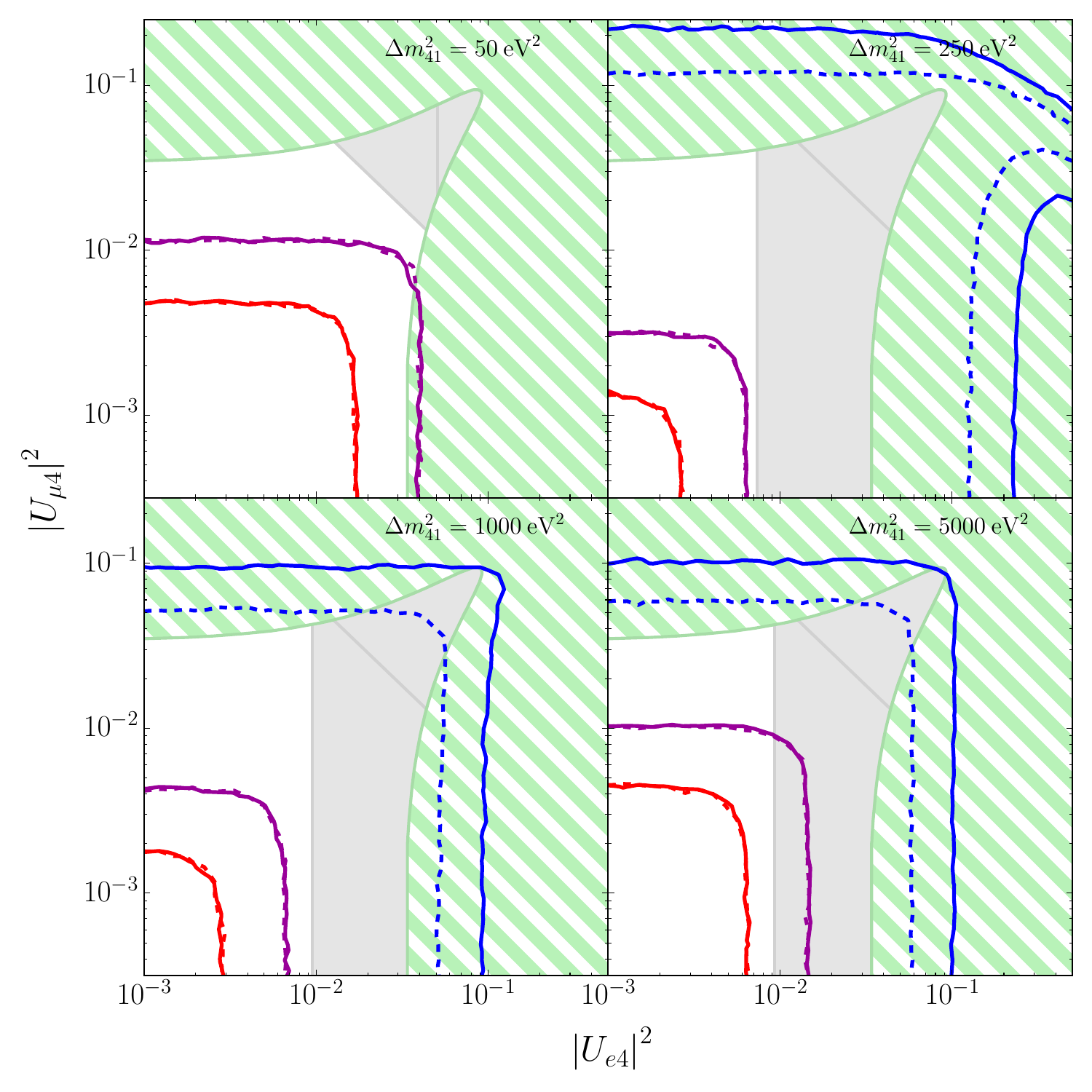}
    \caption{Expected sensitivities (90\% CL) of SHiP after a 5-year operation in the $(|U_{e4}|^2,|U_{\mu4}|^2)$ plane for fixed $\Delta m_{41}^2 = 50,\,250,\,1000,\,5000\,\mathrm{eV}^2$, assuming $|U_{\tau4}|^2=0$. The same line styles and color conventions as in Fig.~\ref{fig:disappearance} are used, while the constraint from Super-Kamiokande~\cite{Super-Kamiokande:2014ndf} is drawn in light green to indicate the additional assumption $\delta_{14}=\delta_{24}=0$. 
    Each grey region shows the constraints from different parameter spaces: regions with vertical boundaries show the limit on $|U_{e4}|^2$ from tritium beta decay experiments~\cite{Belesev:2013cba,KATRIN:2022ith,KATRIN:2022spi}, and regions with diagonal boundaries show the limit on $\sin^22\theta_{e\mu}$ from MicroBooNE~\cite{MicroBooNE:2022sdp}.}
    \label{fig:emu}
\end{figure}

\begin{figure}
    \centering
    \includegraphics[width=.99\textwidth]{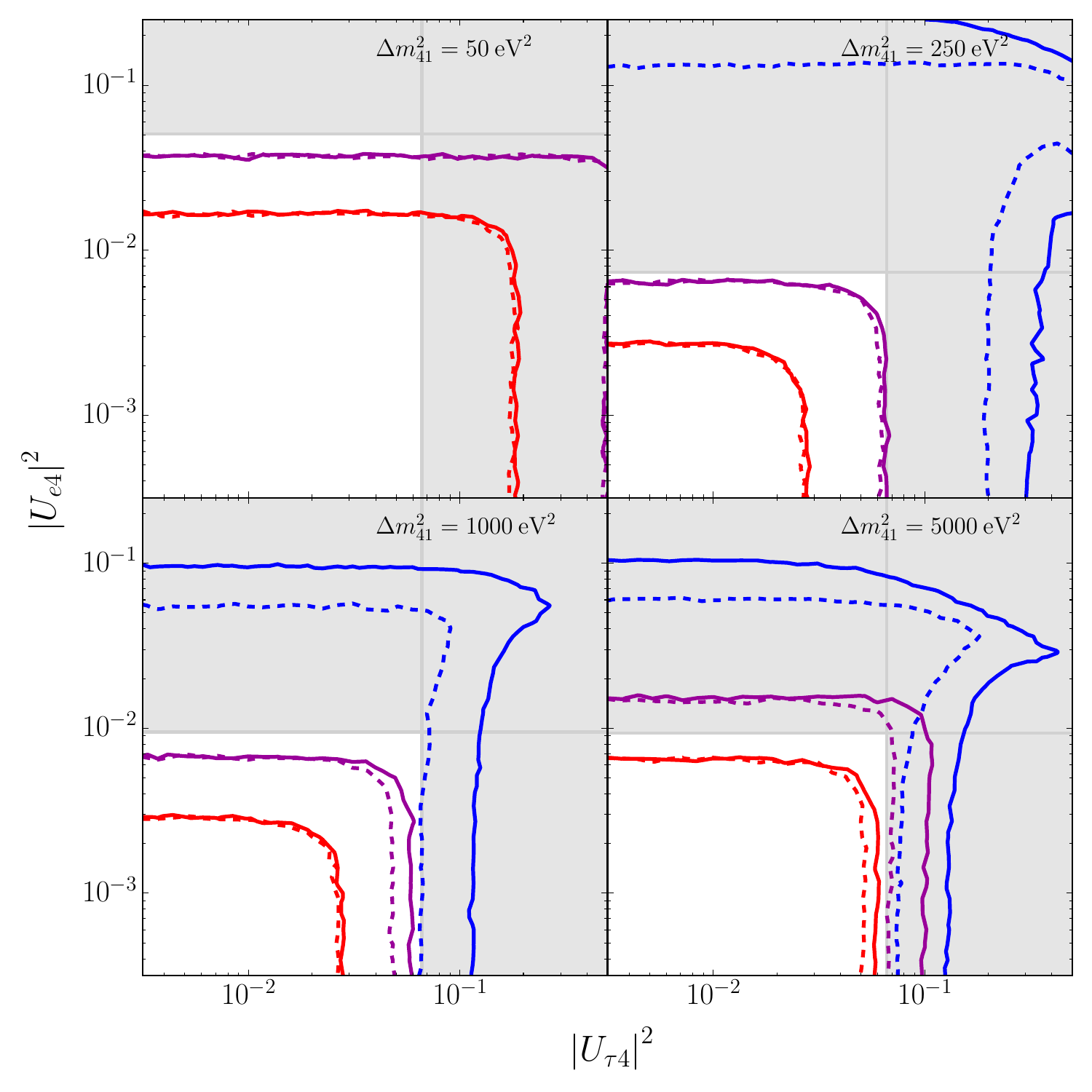}
    \caption{Same as Fig.~\ref{fig:emu}, but in the $(|U_{\tau 4}|^2,|U_{e 4}|^2)$ plane assuming $|U_{\mu4}|^2=0$. The limit on $|U_{e4}|^2$ is from tritium beta decay experiments~\cite{Belesev:2013cba,KATRIN:2022ith,KATRIN:2022spi} and Super-Kamiokande~\cite{Super-Kamiokande:2014ndf}, the limit on $|U_{\tau4}|^2$ is from IceCube-DeepCore~\cite{IceCube:2024dlz}.}
    \label{fig:etau}
\end{figure}

\begin{figure}
    \centering
    \includegraphics[width=.99\textwidth]{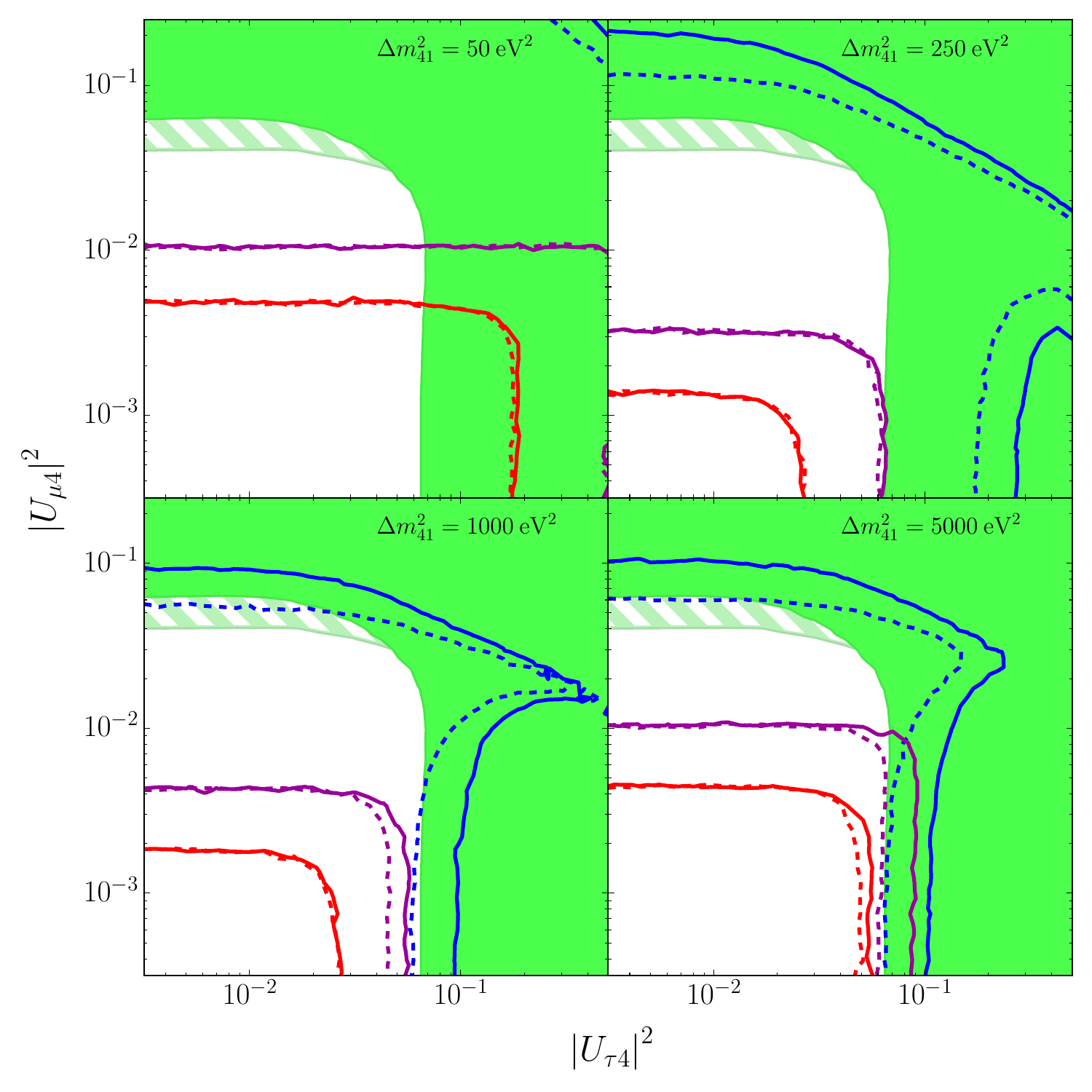}
    \caption{Same as Fig.~\ref{fig:emu}, but in the $(|U_{\tau4}|^2,|U_{\mu4}|^2)$ plane assuming $|U_{e4}|^2=0$. IceCube-DeepCore~\cite{IceCube:2024dlz} is shown in green, while the constraint from Super-Kamiokande~\cite{Super-Kamiokande:2014ndf} is drawn as light green hatched region to indicate the additional assumption $\delta_{24}=0$.}
    \label{fig:mutau}
\end{figure}

Figures~\ref{fig:emu}, \ref{fig:etau}, and \ref{fig:mutau} show the expected sensitivities after a 5-year operation in the $(|U_{\alpha4}|^2,|U_{\beta4}|^2)$ plane for fixed $\Delta m_{41}^2 = 50,\,250,\,1000,\,5000\,\mathrm{eV}^2$.
As before, sensitivities are drawn using dashed (for $\sigma_\mathrm{norm}=10\%$) or solid (for $\sigma_\mathrm{norm}=20\%$) lines with the color coding according to $R_\mathrm{F/N}$.
We draw the constraints from IceCube-DeepCore~\cite{IceCube:2024dlz} on the $(|U_{\tau4}|^2,|U_{\mu4}|^2)$ plane as a green, and constraints from Super-Kamiokande~\cite{Super-Kamiokande:2014ndf} on $(|U_{e4}|^2,|U_{\mu4}|^2)$ and $(|U_{\tau4}|^2,|U_{\mu4}|^2)$ planes are drawn as a light green hatched region to indicate the additional assumption $\delta_{14}=\delta_{24}=0$.
Limits on $|U_{e4}|^2$ at Figs.~\ref{fig:emu} and~\ref{fig:etau} are from tritium beta decay experiments~\cite{Belesev:2013cba,KATRIN:2022ith,KATRIN:2022spi}, and the limit on $|U_{\tau4}|^2$ at Fig.~\ref{fig:etau} comes from IceCube-DeepCore~\cite{IceCube:2024dlz} where the limit is unchanged on varying $|U_{e4}|^2$ as the contribution of electron neutrino CC event is negligible.
Finally, the neutrino appearance limit from MicroBooNE~\cite{MicroBooNE:2022sdp} is shown as diagonal grey regions in Fig.~\ref{fig:emu}, while other neutrino appearance limits are not shown because they could be affected by cancellation effects from neutrino disappearance.

A distinct feature of NSND-only cases is that kinks or discontinuities are shown on the shape of sensitivities, due to a cancellation between appearance and disappearance as we mentioned on Fig.~\ref{fig:binned}. 
Although reducing $\sigma_\mathrm{norm}$ from $20\%$ to $10\%$ improves the sensitivity by approximately a factor of 2, discontinuities and kinks remains.

With FSND, the sensitivities become independent of $\sigma_\mathrm{norm}$ which we already saw in Fig.~\ref{fig:disappearance}. 
The kinks or discontinuities in the sensitivities disappear, which suggests that cross-checks with multiple baselines could distinguish between the cancellation in the 3+1 model and a null result under the three-flavor model.
The enhancement factors due to FSND are similar to those in the single-flavor cases on Fig.~\ref{fig:disappearance}, with approximately a factor of 10 improvement for $|U_{e4}|^2$ and $|U_{\mu4}|^2$, and a factor of 2 improvement for $|U_{\tau4}|^2$ at $\Delta m_{41}^2=1000\,\mathrm{eV}^2$. 
Increasing $R_\mathrm{F/N}$ from $10\%$ to $100\%$ further enhances the sensitivities by roughly 2.3 times.

When compared with existing constraints, only a small portion of the unprobed region on the $(|U_{\tau4}|^2,|U_{\mu4}|^2)$ plane can be constrained in the NSND-only case with $\sigma_\mathrm{norm}=10\%$ at $\Delta m_{41}^2=1000\,\mathrm{eV}^2$, whereas the single-flavor scenario in Fig.~\ref{fig:disappearance} probes the $|U_{\tau4}|^2\gtrsim4\times10^{-2}$ region under the same conditions. 
In contrast, the FSND-enhanced sensitivities can reach $|U_{\alpha4}|^2\lesssim10^{-2}$ for the electron and muon flavors, probing parameter space that are unprobed by past experiments.

%%%%%%%%%%%%%%%%%%%%%%%%%%%%%%%
\section{Conclusion}
\label{sec:conclusion}
%%%%%%%%%%%%%%%%%%%%%%%%%%%%%%%

In this paper, we have investigated the potential of SHiP as a prober of the 3+1 model by estimating sensitivities on various perspectives.
To find sensitivities, we used the FC method and a parametric bootstrap for nuisance parameters, while nuisance parameters and the likelihood of their point estimates are given by hand.
As a successive study of Ref.~\cite{Choi:2024ips}, we also evaluate the effect of additional baseline on the sensitivities of SHiP.

Our results show that on NSND-only cases, SHiP could probe $|U_{\alpha4}|^2\gtrsim0.1$ regions near $\Delta m_{41}^2\sim10^3\,\mathrm{eV}^2$ if $\sigma_\mathrm{norm} = 20\%$.
In two-flavor mixing scenarios kinks and discontinuities of sensitivities are generated due to a cancellation between appearance and disappearance.
Reducing $\sigma_\mathrm{norm}$ from 20\% to 10\% improves the sensitivity by approximately a factor of 2 while reducing $\sigma_\mathrm{norm}$ could not remove any kinks and disappearance in two flavor mixing scenarios.
The inclusion of FSND further enhance the sensitivities, with approximately a factor of 10 for electron and muon flavors, while a factor of 2 to 3 for tau flavors depending on $\sigma_\mathrm{norm}$.
Also, increasing $R_\mathrm{F/N}$ from $10\%$ to $100\%$ further enhances the sensitivity by a factor of 2.3 in all flavors in regions where neutrino oscillations are not averaged.

Compared with existing constraints, the NSND-only configuration does not exceed previous bounds, except for tau flavor mixing when $\sigma_\mathrm{norm}=10\%$.
However, the inclusion of FSND significantly improves the sensitivity. 
Near $\Delta m_{41}^2 \sim 10^3\,\mathrm{eV}^2$, with $R_\mathrm{F/N}=100\%$, the sensitivity for the muon flavor mixing parameter is a factor of 10 stronger than the current limit, and for the tau flavor, the sensitivity is roughly a factor of 7 stronger. 
For the electron flavor, however, the SHiP sensitivity only surpasses the tritium beta decay constraints by a factor of 3.

Taken together, we show that the multi baseline approach gives constraints that are independent with nuisance parameters under the ideal condition, even though sensitivities of NSND can be changed depending on the uncertainty of nuisance parameters.
A cancellation between appearance and disappearance is also ineffective under the dual baseline scenarios as we have seen on Section~\ref{subsec:two}.
Such benefits could help probing sterile neutrinos with more complicated models.

\acknowledgments
The authors were supported by the National Research Foundation of Korea (NRF) grant funded by the Korea government (MEST) 
2021R1A2C2011003, 2021R1F1A1061717, RS-2022-NR069214 and RS-2025-00555834.

%\paragraph{Note added.} This is also a good position for notes added after the paper has been written.

% Bibliography

%% [A] Recommended: using JHEP.bst file
\bibliographystyle{JHEP}
\bibliography{biblio.bib}

\providecommand{\href}[2]{#2}\begingroup\raggedright\begin{thebibliography}{10}

\bibitem{ParticleDataGroup:2024cfk}
{\scshape Particle Data Group} collaboration, \emph{{Review of particle
  physics}}, \href{https://doi.org/10.1103/PhysRevD.110.030001}{\emph{Phys.
  Rev. D} {\bfseries 110} (2024) 030001}.

\bibitem{Kamiokande-II:1991pyu}
{\scshape Kamiokande-II} collaboration, \emph{{Real time, directional
  measurement of B-8 solar neutrinos in the Kamiokande-II detector}},
  \href{https://doi.org/10.1103/PhysRevD.44.2241}{\emph{Phys. Rev. D}
  {\bfseries 44} (1991) 2241}.

\bibitem{GALLEX:1996xbd}
{\scshape GALLEX} collaboration, \emph{{GALLEX solar neutrino observations:
  Results for GALLEX III.}},
  \href{https://doi.org/10.1016/S0370-2693(96)01121-5}{\emph{Phys. Lett. B}
  {\bfseries 388} (1996) 384}.

\bibitem{Super-Kamiokande:2001ljr}
{\scshape Super-Kamiokande} collaboration, \emph{{Solar B-8 and hep neutrino
  measurements from 1258 days of Super-Kamiokande data}},
  \href{https://doi.org/10.1103/PhysRevLett.86.5651}{\emph{Phys. Rev. Lett.}
  {\bfseries 86} (2001) 5651}
  [\href{https://arxiv.org/abs/hep-ex/0103032}{{\ttfamily hep-ex/0103032}}].

\bibitem{SNO:2002tuh}
{\scshape SNO} collaboration, \emph{{Direct evidence for neutrino flavor
  transformation from neutral current interactions in the Sudbury Neutrino
  Observatory}},
  \href{https://doi.org/10.1103/PhysRevLett.89.011301}{\emph{Phys. Rev. Lett.}
  {\bfseries 89} (2002) 011301}
  [\href{https://arxiv.org/abs/nucl-ex/0204008}{{\ttfamily nucl-ex/0204008}}].

\bibitem{SAGE:2009eeu}
{\scshape SAGE} collaboration, \emph{{Measurement of the solar neutrino capture
  rate with gallium metal. III: Results for the 2002--2007 data-taking
  period}}, \href{https://doi.org/10.1103/PhysRevC.80.015807}{\emph{Phys. Rev.
  C} {\bfseries 80} (2009) 015807}
  [\href{https://arxiv.org/abs/0901.2200}{{\ttfamily 0901.2200}}].

\bibitem{Borexino:2013zhu}
{\scshape Borexino} collaboration, \emph{{Final results of Borexino Phase-I on
  low energy solar neutrino spectroscopy}},
  \href{https://doi.org/10.1103/PhysRevD.89.112007}{\emph{Phys. Rev. D}
  {\bfseries 89} (2014) 112007}
  [\href{https://arxiv.org/abs/1308.0443}{{\ttfamily 1308.0443}}].

\bibitem{Super-Kamiokande:1998kpq}
{\scshape Super-Kamiokande} collaboration, \emph{{Evidence for oscillation of
  atmospheric neutrinos}},
  \href{https://doi.org/10.1103/PhysRevLett.81.1562}{\emph{Phys. Rev. Lett.}
  {\bfseries 81} (1998) 1562}
  [\href{https://arxiv.org/abs/hep-ex/9807003}{{\ttfamily hep-ex/9807003}}].

\bibitem{ANTARES:2012tem}
{\scshape ANTARES} collaboration, \emph{{Measurement of Atmospheric Neutrino
  Oscillations with the ANTARES Neutrino Telescope}},
  \href{https://doi.org/10.1016/j.physletb.2012.07.002}{\emph{Phys. Lett. B}
  {\bfseries 714} (2012) 224}
  [\href{https://arxiv.org/abs/1206.0645}{{\ttfamily 1206.0645}}].

\bibitem{IceCube:2013pav}
{\scshape IceCube} collaboration, \emph{{Measurement of Atmospheric Neutrino
  Oscillations with IceCube}},
  \href{https://doi.org/10.1103/PhysRevLett.111.081801}{\emph{Phys. Rev. Lett.}
  {\bfseries 111} (2013) 081801}
  [\href{https://arxiv.org/abs/1305.3909}{{\ttfamily 1305.3909}}].

\bibitem{MINOS:2008kxu}
{\scshape MINOS} collaboration, \emph{{Measurement of Neutrino Oscillations
  with the MINOS Detectors in the NuMI Beam}},
  \href{https://doi.org/10.1103/PhysRevLett.101.131802}{\emph{Phys. Rev. Lett.}
  {\bfseries 101} (2008) 131802}
  [\href{https://arxiv.org/abs/0806.2237}{{\ttfamily 0806.2237}}].

\bibitem{T2K:2013bzi}
{\scshape T2K} collaboration, \emph{{Measurement of Neutrino Oscillation
  Parameters from Muon Neutrino Disappearance with an Off-axis Beam}},
  \href{https://doi.org/10.1103/PhysRevLett.111.211803}{\emph{Phys. Rev. Lett.}
  {\bfseries 111} (2013) 211803}
  [\href{https://arxiv.org/abs/1308.0465}{{\ttfamily 1308.0465}}].

\bibitem{NOvA:2016vij}
{\scshape NOvA} collaboration, \emph{{First measurement of muon-neutrino
  disappearance in NOvA}},
  \href{https://doi.org/10.1103/PhysRevD.93.051104}{\emph{Phys. Rev. D}
  {\bfseries 93} (2016) 051104}
  [\href{https://arxiv.org/abs/1601.05037}{{\ttfamily 1601.05037}}].

\bibitem{OPERA:2018nar}
{\scshape OPERA} collaboration, \emph{{Final Results of the OPERA Experiment on
  $\nu_\tau$ Appearance in the CNGS Neutrino Beam}},
  \href{https://doi.org/10.1103/PhysRevLett.120.211801}{\emph{Phys. Rev. Lett.}
  {\bfseries 120} (2018) 211801}
  [\href{https://arxiv.org/abs/1804.04912}{{\ttfamily 1804.04912}}].

\bibitem{IceCube:2024pky}
{\scshape IceCube} collaboration, \emph{{Exploration of mass splitting and
  muon/tau mixing parameters for an eV-scale sterile neutrino with IceCube}},
  \href{https://doi.org/10.1016/j.physletb.2024.139077}{\emph{Phys. Lett. B}
  {\bfseries 858} (2024) 139077}
  [\href{https://arxiv.org/abs/2406.00905}{{\ttfamily 2406.00905}}].

\bibitem{LSND:2001aii}
{\scshape LSND} collaboration, \emph{{Evidence for neutrino oscillations from
  the observation of $\bar{\nu}_e$ appearance in a $\bar{\nu}_\mu$ beam}},
  \href{https://doi.org/10.1103/PhysRevD.64.112007}{\emph{Phys. Rev. D}
  {\bfseries 64} (2001) 112007}
  [\href{https://arxiv.org/abs/hep-ex/0104049}{{\ttfamily hep-ex/0104049}}].

\bibitem{MiniBooNE:2020pnu}
{\scshape MiniBooNE} collaboration, \emph{{Updated MiniBooNE neutrino
  oscillation results with increased data and new background studies}},
  \href{https://doi.org/10.1103/PhysRevD.103.052002}{\emph{Phys. Rev. D}
  {\bfseries 103} (2021) 052002}
  [\href{https://arxiv.org/abs/2006.16883}{{\ttfamily 2006.16883}}].

\bibitem{Mention:2011rk}
G.~Mention, M.~Fechner, T.~Lasserre, T.A.~Mueller, D.~Lhuillier, M.~Cribier
  et~al., \emph{{The Reactor Antineutrino Anomaly}},
  \href{https://doi.org/10.1103/PhysRevD.83.073006}{\emph{Phys. Rev. D}
  {\bfseries 83} (2011) 073006}
  [\href{https://arxiv.org/abs/1101.2755}{{\ttfamily 1101.2755}}].

\bibitem{Elliott:2023cvh}
S.R.~Elliott, V.~Gavrin and W.~Haxton, \emph{{The gallium anomaly}},
  \href{https://doi.org/10.1016/j.ppnp.2023.104082}{\emph{Prog. Part. Nucl.
  Phys.} {\bfseries 134} (2024) 104082}
  [\href{https://arxiv.org/abs/2306.03299}{{\ttfamily 2306.03299}}].

\bibitem{Minkowski:1977sc}
P.~Minkowski, \emph{{$\mu \to e\gamma$ at a Rate of One Out of $10^{9}$ Muon
  Decays?}}, \href{https://doi.org/10.1016/0370-2693(77)90435-X}{\emph{Phys.
  Lett. B} {\bfseries 67} (1977) 421}.

\bibitem{Ramond:1979py}
P.~Ramond, \emph{{The Family Group in Grand Unified Theories}},  in
  \emph{{International Symposium on Fundamentals of Quantum Theory and Quantum
  Field Theory}}, 2, 1979
  [\href{https://arxiv.org/abs/hep-ph/9809459}{{\ttfamily hep-ph/9809459}}].

\bibitem{Gell-Mann:1979vob}
M.~Gell-Mann, P.~Ramond and R.~Slansky, \emph{{Complex Spinors and Unified
  Theories}}, {\emph{Conf. Proc. C} {\bfseries 790927} (1979) 315}
  [\href{https://arxiv.org/abs/1306.4669}{{\ttfamily 1306.4669}}].

\bibitem{Yanagida:1979as}
T.~Yanagida, \emph{{Horizontal gauge symmetry and masses of neutrinos}},
  {\emph{Conf. Proc. C} {\bfseries 7902131} (1979) 95}.

\bibitem{Mohapatra:1979ia}
R.N.~Mohapatra and G.~Senjanovic, \emph{{Neutrino Mass and Spontaneous Parity
  Nonconservation}},
  \href{https://doi.org/10.1103/PhysRevLett.44.912}{\emph{Phys. Rev. Lett.}
  {\bfseries 44} (1980) 912}.

\bibitem{Boyarsky:2018tvu}
A.~Boyarsky, M.~Drewes, T.~Lasserre, S.~Mertens and O.~Ruchayskiy,
  \emph{{Sterile neutrino Dark Matter}},
  \href{https://doi.org/10.1016/j.ppnp.2018.07.004}{\emph{Prog. Part. Nucl.
  Phys.} {\bfseries 104} (2019) 1}
  [\href{https://arxiv.org/abs/1807.07938}{{\ttfamily 1807.07938}}].

\bibitem{Asaka:2005pn}
T.~Asaka and M.~Shaposhnikov, \emph{{The $\nu$MSM, dark matter and baryon
  asymmetry of the universe}},
  \href{https://doi.org/10.1016/j.physletb.2005.06.020}{\emph{Phys. Lett. B}
  {\bfseries 620} (2005) 17}
  [\href{https://arxiv.org/abs/hep-ph/0505013}{{\ttfamily hep-ph/0505013}}].

\bibitem{Acero:2022wqg}
M.A.~Acero et~al., \emph{{White Paper on Light Sterile Neutrino Searches and
  Related Phenomenology}},  \href{https://arxiv.org/abs/2203.07323}{{\ttfamily
  2203.07323}}.

\bibitem{Super-Kamiokande:2014ndf}
{\scshape Super-Kamiokande} collaboration, \emph{{Limits on sterile neutrino
  mixing using atmospheric neutrinos in Super-Kamiokande}},
  \href{https://doi.org/10.1103/PhysRevD.91.052019}{\emph{Phys. Rev. D}
  {\bfseries 91} (2015) 052019}
  [\href{https://arxiv.org/abs/1410.2008}{{\ttfamily 1410.2008}}].

\bibitem{NOvA:2017geg}
{\scshape NOvA} collaboration, \emph{{Search for active-sterile neutrino mixing
  using neutral-current interactions in NOvA}},
  \href{https://doi.org/10.1103/PhysRevD.96.072006}{\emph{Phys. Rev. D}
  {\bfseries 96} (2017) 072006}
  [\href{https://arxiv.org/abs/1706.04592}{{\ttfamily 1706.04592}}].

\bibitem{MINOS:2017cae}
{\scshape MINOS+} collaboration, \emph{{Search for sterile neutrinos in MINOS
  and MINOS+ using a two-detector fit}},
  \href{https://doi.org/10.1103/PhysRevLett.122.091803}{\emph{Phys. Rev. Lett.}
  {\bfseries 122} (2019) 091803}
  [\href{https://arxiv.org/abs/1710.06488}{{\ttfamily 1710.06488}}].

\bibitem{T2K:2019efw}
{\scshape T2K} collaboration, \emph{{Search for light sterile neutrinos with
  the T2K far detector Super-Kamiokande at a baseline of 295 km}},
  \href{https://doi.org/10.1103/PhysRevD.99.071103}{\emph{Phys. Rev. D}
  {\bfseries 99} (2019) 071103}
  [\href{https://arxiv.org/abs/1902.06529}{{\ttfamily 1902.06529}}].

\bibitem{OPERA:2019kzo}
{\scshape OPERA} collaboration, \emph{{Final results on neutrino oscillation
  parameters from the OPERA experiment in the CNGS beam}},
  \href{https://doi.org/10.1103/PhysRevD.100.051301}{\emph{Phys. Rev. D}
  {\bfseries 100} (2019) 051301}
  [\href{https://arxiv.org/abs/1904.05686}{{\ttfamily 1904.05686}}].

\bibitem{IceCube:2024dlz}
{\scshape IceCube} collaboration, \emph{{Search for a light sterile neutrino
  with 7.5~years of IceCube DeepCore data}},
  \href{https://doi.org/10.1103/PhysRevD.110.072007}{\emph{Phys. Rev. D}
  {\bfseries 110} (2024) 072007}
  [\href{https://arxiv.org/abs/2407.01314}{{\ttfamily 2407.01314}}].

\bibitem{DANSS:2018fnn}
{\scshape DANSS} collaboration, \emph{{Search for sterile neutrinos at the
  DANSS experiment}},
  \href{https://doi.org/10.1016/j.physletb.2018.10.038}{\emph{Phys. Lett. B}
  {\bfseries 787} (2018) 56}
  [\href{https://arxiv.org/abs/1804.04046}{{\ttfamily 1804.04046}}].

\bibitem{NEOS:2016wee}
{\scshape NEOS} collaboration, \emph{{Sterile Neutrino Search at the NEOS
  Experiment}},
  \href{https://doi.org/10.1103/PhysRevLett.118.121802}{\emph{Phys. Rev. Lett.}
  {\bfseries 118} (2017) 121802}
  [\href{https://arxiv.org/abs/1610.05134}{{\ttfamily 1610.05134}}].

\bibitem{HOUMMADA1995449}
A.~Hoummada, S.~{Lazrak Mikou}, M.~Avenier, G.~Bagieu, J.~Cavaignac and
  D.~{Holm Koang}, \emph{Neutrino oscillations i.l.l. experiment reanalysis},
  \href{https://doi.org/https://doi.org/10.1016/0969-8043(95)00048-8}{\emph{Applied
  Radiation and Isotopes} {\bfseries 46} (1995) 449}.

\bibitem{PROSPECT:2018dtt}
{\scshape PROSPECT} collaboration, \emph{{First search for short-baseline
  neutrino oscillations at HFIR with PROSPECT}},
  \href{https://doi.org/10.1103/PhysRevLett.121.251802}{\emph{Phys. Rev. Lett.}
  {\bfseries 121} (2018) 251802}
  [\href{https://arxiv.org/abs/1806.02784}{{\ttfamily 1806.02784}}].

\bibitem{STEREO:2018rfh}
{\scshape STEREO} collaboration, \emph{{Sterile Neutrino Constraints from the
  STEREO Experiment with 66 Days of Reactor-On Data}},
  \href{https://doi.org/10.1103/PhysRevLett.121.161801}{\emph{Phys. Rev. Lett.}
  {\bfseries 121} (2018) 161801}
  [\href{https://arxiv.org/abs/1806.02096}{{\ttfamily 1806.02096}}].

\bibitem{RENO:2020hva}
{\scshape RENO, NEOS} collaboration, \emph{{Search for sterile neutrino
  oscillations using RENO and NEOS data}},
  \href{https://doi.org/10.1103/PhysRevD.105.L111101}{\emph{Phys. Rev. D}
  {\bfseries 105} (2022) L111101}
  [\href{https://arxiv.org/abs/2011.00896}{{\ttfamily 2011.00896}}].

\bibitem{RENO:2020uip}
{\scshape RENO} collaboration, \emph{{Search for Sub-eV Sterile Neutrinos at
  RENO}}, \href{https://doi.org/10.1103/PhysRevLett.125.191801}{\emph{Phys.
  Rev. Lett.} {\bfseries 125} (2020) 191801}
  [\href{https://arxiv.org/abs/2006.07782}{{\ttfamily 2006.07782}}].

\bibitem{DayaBay:2024nip}
{\scshape Daya Bay} collaboration, \emph{{Search for a Sub-eV Sterile Neutrino
  using Daya Bay\textquoteright{}s Full Dataset}},
  \href{https://doi.org/10.1103/PhysRevLett.133.051801}{\emph{Phys. Rev. Lett.}
  {\bfseries 133} (2024) 051801}
  [\href{https://arxiv.org/abs/2404.01687}{{\ttfamily 2404.01687}}].

\bibitem{Kraus:2012he}
C.~Kraus, A.~Singer, K.~Valerius and C.~Weinheimer, \emph{{Limit on sterile
  neutrino contribution from the Mainz Neutrino Mass Experiment}},
  \href{https://doi.org/10.1140/epjc/s10052-013-2323-z}{\emph{Eur. Phys. J. C}
  {\bfseries 73} (2013) 2323}
  [\href{https://arxiv.org/abs/1210.4194}{{\ttfamily 1210.4194}}].

\bibitem{Belesev:2012hx}
A.I.~Belesev, A.I.~Berlev, E.V.~Geraskin, A.A.~Golubev, N.A.~Likhovid,
  A.A.~Nozik et~al., \emph{{An upper limit on additional neutrino mass
  eigenstate in 2 to 100 eV region from 'Troitsk nu-mass' data}},
  \href{https://doi.org/10.1134/S0021364013020033}{\emph{JETP Lett.} {\bfseries
  97} (2013) 67} [\href{https://arxiv.org/abs/1211.7193}{{\ttfamily
  1211.7193}}].

\bibitem{Belesev:2013cba}
A.I.~Belesev, A.I.~Berlev, E.V.~Geraskin, A.A.~Golubev, N.A.~Likhovid,
  A.A.~Nozik et~al., \emph{{The search for an additional neutrino mass
  eigenstate in the 2 to 100 eV region from Troitsk nu-mass data: a detailed
  analysis}}, \href{https://doi.org/10.1088/0954-3899/41/1/015001}{\emph{J.
  Phys. G} {\bfseries 41} (2014) 015001}
  [\href{https://arxiv.org/abs/1307.5687}{{\ttfamily 1307.5687}}].

\bibitem{Abdurashitov:2017kka}
J.N.~Abdurashitov et~al., \emph{{First measeurements in search for keV-sterile
  neutrino in tritium beta-decay by Troitsk nu-mass experiment}},
  \href{https://doi.org/10.1134/S0021364017120013}{\emph{Pisma Zh. Eksp. Teor.
  Fiz.} {\bfseries 105} (2017) 723}
  [\href{https://arxiv.org/abs/1703.10779}{{\ttfamily 1703.10779}}].

\bibitem{KATRIN:2022ith}
{\scshape KATRIN} collaboration, \emph{{Improved eV-scale sterile-neutrino
  constraints from the second KATRIN measurement campaign}},
  \href{https://doi.org/10.1103/PhysRevD.105.072004}{\emph{Phys. Rev. D}
  {\bfseries 105} (2022) 072004}
  [\href{https://arxiv.org/abs/2201.11593}{{\ttfamily 2201.11593}}].

\bibitem{KATRIN:2022spi}
{\scshape KATRIN} collaboration, \emph{{Search for keV-scale sterile neutrinos
  with the first KATRIN data}},
  \href{https://doi.org/10.1140/epjc/s10052-023-11818-y}{\emph{Eur. Phys. J. C}
  {\bfseries 83} (2023) 763}
  [\href{https://arxiv.org/abs/2207.06337}{{\ttfamily 2207.06337}}].

\bibitem{DYDAK1984281}
F.~Dydak, G.~Feldman, C.~Guyot, J.~Merlo, H.-J.~Meyer, J.~Rothberg et~al.,
  \emph{A search for $\nu_\mu$ oscillations in the $\delta m^2$ range
  $0.3-90\,\mathrm{eV}^2$},
  \href{https://doi.org/https://doi.org/10.1016/0370-2693(84)90688-9}{\emph{Physics
  Letters B} {\bfseries 134} (1984) 281}.

\bibitem{PhysRevLett.52.1384}
I.E.~Stockdale, A.~Bodek, F.~Borcherding, N.~Giokaris, K.~Lang, D.~Garfinkle
  et~al., \emph{Limits on muon-neutrino oscillations in the mass range
  $30<\delta m^{2}<1000$ $\mathrm{eV}^{2}$/${\mathit{c}}^{4}$},
  \href{https://doi.org/10.1103/PhysRevLett.52.1384}{\emph{Phys. Rev. Lett.}
  {\bfseries 52} (1984) 1384}.

\bibitem{CCFRNuTeV:1998gjj}
{\scshape CCFR/NuTeV} collaboration, \emph{{A High statistics search for
  neutrino(e) (anti-neutrino(e)) $\longrightarrow$ neutrino(tau)
  (anti-neutrino(tau)) oscillations}},
  \href{https://doi.org/10.1103/PhysRevD.59.031101}{\emph{Phys. Rev. D}
  {\bfseries 59} (1999) 031101}
  [\href{https://arxiv.org/abs/hep-ex/9809023}{{\ttfamily hep-ex/9809023}}].

\bibitem{NOMAD:2001xxt}
{\scshape NOMAD} collaboration, \emph{{Final NOMAD results on muon-neutrino
  $\longrightarrow$ tau-neutrino and electron-neutrino $\longrightarrow$
  tau-neutrino oscillations including a new search for tau-neutrino appearance
  using hadronic tau decays}},
  \href{https://doi.org/10.1016/S0550-3213(01)00339-X}{\emph{Nucl. Phys. B}
  {\bfseries 611} (2001) 3}
  [\href{https://arxiv.org/abs/hep-ex/0106102}{{\ttfamily hep-ex/0106102}}].

\bibitem{NuTeV:2002daf}
{\scshape NuTeV} collaboration, \emph{{A Search for $\nu_{\mu} \to \nu_e$ and
  $\bar{\nu}_{\mu} \to \bar \nu_e$ Oscillations at NuTeV}},
  \href{https://doi.org/10.1103/PhysRevLett.89.011804}{\emph{Phys. Rev. Lett.}
  {\bfseries 89} (2002) 011804}
  [\href{https://arxiv.org/abs/hep-ex/0203018}{{\ttfamily hep-ex/0203018}}].

\bibitem{KARMEN:2002zcm}
{\scshape KARMEN} collaboration, \emph{{Upper limits for neutrino oscillations
  muon-anti-neutrino $\longrightarrow$ electron-anti-neutrino from muon decay
  at rest}}, \href{https://doi.org/10.1103/PhysRevD.65.112001}{\emph{Phys. Rev.
  D} {\bfseries 65} (2002) 112001}
  [\href{https://arxiv.org/abs/hep-ex/0203021}{{\ttfamily hep-ex/0203021}}].

\bibitem{NOMAD:2003mqg}
{\scshape NOMAD} collaboration, \emph{{Search for nu(mu) $\longrightarrow$
  nu(e) oscillations in the NOMAD experiment}},
  \href{https://doi.org/10.1016/j.physletb.2003.07.029}{\emph{Phys. Lett. B}
  {\bfseries 570} (2003) 19}
  [\href{https://arxiv.org/abs/hep-ex/0306037}{{\ttfamily hep-ex/0306037}}].

\bibitem{CHORUS:2007wlo}
{\scshape CHORUS} collaboration, \emph{{Final results on nu(mu)
  $\longrightarrow$ nu(tau) oscillation from the CHORUS experiment}},
  \href{https://doi.org/10.1016/j.nuclphysb.2007.10.023}{\emph{Nucl. Phys. B}
  {\bfseries 793} (2008) 326}
  [\href{https://arxiv.org/abs/0710.3361}{{\ttfamily 0710.3361}}].

\bibitem{MiniBooNE:2012meu}
{\scshape MiniBooNE, SciBooNE} collaboration, \emph{{Dual baseline search for
  muon antineutrino disappearance at $0.1\,\mathrm{eV}^2 < {\Delta}m^2 <
  100\,\mathrm{eV}^2$}},
  \href{https://doi.org/10.1103/PhysRevD.86.052009}{\emph{Phys. Rev. D}
  {\bfseries 86} (2012) 052009}
  [\href{https://arxiv.org/abs/1208.0322}{{\ttfamily 1208.0322}}].

\bibitem{MicroBooNE:2022sdp}
{\scshape MicroBooNE} collaboration, \emph{{First Constraints on Light Sterile
  Neutrino Oscillations from Combined Appearance and Disappearance Searches
  with the MicroBooNE Detector}},
  \href{https://doi.org/10.1103/PhysRevLett.130.011801}{\emph{Phys. Rev. Lett.}
  {\bfseries 130} (2023) 011801}
  [\href{https://arxiv.org/abs/2210.10216}{{\ttfamily 2210.10216}}].

\bibitem{Bai:2020ukz}
W.~Bai, M.~Diwan, M.V.~Garzelli, Y.S.~Jeong and M.H.~Reno, \emph{{Far-forward
  neutrinos at the Large Hadron Collider}},
  \href{https://doi.org/10.1007/JHEP06(2020)032}{\emph{JHEP} {\bfseries 06}
  (2020) 032} [\href{https://arxiv.org/abs/2002.03012}{{\ttfamily
  2002.03012}}].

\bibitem{FASER:2019dxq}
{\scshape FASER} collaboration, \emph{{Detecting and Studying High-Energy
  Collider Neutrinos with FASER at the LHC}},
  \href{https://doi.org/10.1140/epjc/s10052-020-7631-5}{\emph{Eur. Phys. J. C}
  {\bfseries 80} (2020) 61} [\href{https://arxiv.org/abs/1908.02310}{{\ttfamily
  1908.02310}}].

\bibitem{SHiP:2015vad}
{\scshape SHiP} collaboration, \emph{{A facility to Search for Hidden Particles
  (SHiP) at the CERN SPS}},  \href{https://arxiv.org/abs/1504.04956}{{\ttfamily
  1504.04956}}.

\bibitem{Alekhin:2015byh}
S.~Alekhin et~al., \emph{{A facility to Search for Hidden Particles at the CERN
  SPS: the SHiP physics case}},
  \href{https://doi.org/10.1088/0034-4885/79/12/124201}{\emph{Rept. Prog.
  Phys.} {\bfseries 79} (2016) 124201}
  [\href{https://arxiv.org/abs/1504.04855}{{\ttfamily 1504.04855}}].

\bibitem{Ahdida:2654870}
{\scshape SHiP} collaboration, \emph{{SHiP Experiment - Progress Report}},
  Tech. Rep. \href{https://cds.cern.ch/record/2654870}{CERN-SPSC-2019-010,
  SPSC-SR-248}, CERN, Geneva (2019).

\bibitem{Albanese:2878604}
{\scshape SHiP} collaboration, \emph{{BDF/SHiP at the ECN3 high-intensity beam
  facility}},  Proposal to the CERN SPSC
  \href{https://cds.cern.ch/record/2878604}{CERN-SPSC-2023-033, SPSC-P-369},
  CERN, Geneva (2023).

\bibitem{Choi:2024ips}
K.-Y.~Choi, S.H.~Kim, Y.G.~Kim, K.Y.~Lee, K.S.~Lee, B.D.~Park et~al.,
  \emph{{Probing the mixing between sterile and tau neutrinos in the SHiP
  experiment}}, \href{https://doi.org/10.1007/JHEP06(2024)166}{\emph{JHEP}
  {\bfseries 06} (2024) 166}
  [\href{https://arxiv.org/abs/2403.04191}{{\ttfamily 2403.04191}}].

\bibitem{HARARI1986123}
H.~Harari and M.~Leurer, \emph{Recommending a standard choice of cabibbo angles
  and km phases for any number of generations},
  \href{https://doi.org/https://doi.org/10.1016/0370-2693(86)91268-2}{\emph{Physics
  Letters B} {\bfseries 181} (1986) 123}.

\bibitem{Feldman:1997qc}
G.J.~Feldman and R.D.~Cousins, \emph{{A Unified approach to the classical
  statistical analysis of small signals}},
  \href{https://doi.org/10.1103/PhysRevD.57.3873}{\emph{Phys. Rev. D}
  {\bfseries 57} (1998) 3873}
  [\href{https://arxiv.org/abs/physics/9711021}{{\ttfamily physics/9711021}}].

\bibitem{Demortier2012}
L.~Demortier, \emph{The parametric bootstrap and particle physics},  in
  \emph{Progress on Statistical Issues in Searches: A Conference Involving
  Statistical Issues in Astrophysics, Particle Physics and Photon Science},
  (SLAC National Accelerator Laboratory), June 4--6, 2012,
  \href{https://hep-physics.rockefeller.edu/luc/talks/ParametricBootstrap.pdf}{https://hep-physics.rockefeller.edu/luc/talks/ParametricBootstrap.pdf}.

\bibitem{stuart1999kendalls}
A.~Stuart, K.~Ord and S.~Arnold, \emph{Kendall's Advanced Theory of Statistics,
  Volume 2A: Classical Inference and the Linear Model}, Arnold, London, 6th~ed.
  (1999).

\bibitem{Cousins:2018tiz}
R.D.~Cousins, \emph{{Lectures on Statistics in Theory: Prelude to Statistics in
  Practice}},  \href{https://arxiv.org/abs/1807.05996}{{\ttfamily 1807.05996}}.

\bibitem{Berger01091994}
R.L.~Berger and D.D.B.~and, \emph{P values maximized over a confidence set for
  the nuisance parameter},
  \href{https://doi.org/10.1080/01621459.1994.10476836}{\emph{Journal of the
  American Statistical Association} {\bfseries 89} (1994) 1012}
  [\href{https://arxiv.org/abs/https://doi.org/10.1080/01621459.1994.10476836}{{\ttfamily
  https://doi.org/10.1080/01621459.1994.10476836}}].

\bibitem{COUSINS1992331}
R.D.~Cousins and V.L.~Highland, \emph{Incorporating systematic uncertainties
  into an upper limit},
  \href{https://doi.org/https://doi.org/10.1016/0168-9002(92)90794-5}{\emph{Nuclear
  Instruments and Methods in Physics Research Section A: Accelerators,
  Spectrometers, Detectors and Associated Equipment} {\bfseries 320} (1992)
  331}.

\bibitem{Cranmer:2014lly}
K.~Cranmer, \emph{{Practical Statistics for the LHC}},  in \emph{{2011 European
  School of High-Energy Physics}}, pp.~267--308, 2014,
  \href{https://doi.org/10.5170/CERN-2014-003.267}{DOI}
  [\href{https://arxiv.org/abs/1503.07622}{{\ttfamily 1503.07622}}].

\bibitem{NOvA:2022wnj}
{\scshape NOvA} collaboration, \emph{{The Profiled Feldman-Cousins technique
  for confidence interval construction in the presence of nuisance
  parameters}},  \href{https://arxiv.org/abs/2207.14353}{{\ttfamily
  2207.14353}}.

\bibitem{Buonaura:2268663}
A.~Buonaura, \emph{Study of $\nu_\tau$ Properties with the SHiP Experiment},
  Ph.D. thesis, University of Naples, 2017.

\bibitem{Bai:2018xum}
W.~Bai and M.H.~Reno, \emph{{Prompt neutrinos and intrinsic charm at SHiP}},
  \href{https://doi.org/10.1007/JHEP02(2019)077}{\emph{JHEP} {\bfseries 02}
  (2019) 077} [\href{https://arxiv.org/abs/1807.02746}{{\ttfamily
  1807.02746}}].

\bibitem{DsTau:2019wjb}
{\scshape DsTau} collaboration, \emph{{DsTau: Study of tau neutrino production
  with 400 GeV protons from the CERN-SPS}},
  \href{https://doi.org/10.1007/JHEP01(2020)033}{\emph{JHEP} {\bfseries 01}
  (2020) 033} [\href{https://arxiv.org/abs/1906.03487}{{\ttfamily
  1906.03487}}].

\bibitem{Iuliano:2776128}
A.~Iuliano, \emph{{Event reconstruction and data analysis techniques for the
  SHiP experiment}}, Ph.D. thesis, Università degli Studi di Napoli Federico
  II, 2021.

\bibitem{Feickert:2021ajf}
M.~Feickert and B.~Nachman, \emph{{A Living Review of Machine Learning for
  Particle Physics}},  \href{https://arxiv.org/abs/2102.02770}{{\ttfamily
  2102.02770}}.

\bibitem{NOvA:2024imi}
{\scshape NOvA} collaboration, \emph{{Dual-Baseline Search for
  Active-to-Sterile Neutrino Oscillations in NOvA}},
  \href{https://doi.org/10.1103/PhysRevLett.134.081804}{\emph{Phys. Rev. Lett.}
  {\bfseries 134} (2025) 081804}
  [\href{https://arxiv.org/abs/2409.04553}{{\ttfamily 2409.04553}}].

\end{thebibliography}\endgroup

\end{document}